# Aprendizaje Automatizado para la Identificación de Potenciales Participantes de un Programa Social en Uruguay

Christian Berón Curti

Programa de maestría en Data Science, Massachusetts Institute of Technology - Universidad Tecnológica del Uruguay

cberon@mit.edu / cberon@gmail.com

Rodrigo Vargas Sainz

Programa de maestría en Data Science, Massachusetts Institute of Technology - Universidad Tecnológica del Uruguay

rodrivs@mit.edu / rodrigosj@gmail.com

Yitong Tseo

Massachusetts Institute of Technology

yitongt@mit.edu

**RESUMEN**

Este proyecto de investigación explora la optimización del proceso de selección de familias para participar en el Programa de Acompañamiento Familiar (PAF) de Uruguay Crece Contigo mediante el aprendizaje automático. Se analizó una base de datos anonimizada de 15.436 casos de derivaciones anteriores, centradas en mujeres embarazadas y niños menores de cuatro años. El objetivo principal fue desarrollar un algoritmo predictivo capaz de determinar si una familia cumple con las condiciones para ser aceptada en el programa. La implementación de este modelo busca agilizar el proceso de evaluación y permitir una asignación más eficiente de los recursos, destinando más tiempo del equipo al acompañamiento directo. El estudio abarcó un análisis exhaustivo de los datos y la implementación de diversos modelos de aprendizaje automático, incluyendo Redes Neuronales (NN), XGBoost (XGB), LSTM y modelos ensamblados. Se aplicaron técnicas para manejar el desbalance de clases, como SMOTE y RUS, así como la optimización de umbrales de decisión para mejorar la precisión y el equilibrio de las predicciones. Los resultados obtenidos demuestran el potencial de estas técnicas para una clasificación eficiente de las familias que requieren asistencia.

**CCS CONCEPTS**

- **Applied computing → Health care information systems; Social work.**
- **Information systems → Information extraction; Machine learning.**

## 1. INTRODUCCIÓN

La primera infancia se reconoce como un período crucial, y Uruguay Crece Contigo es un área que busca garantizar los cuidados, el desarrollo y la protección de mujeres embarazadas y familias con niños pequeños. El acceso al programa se realiza mediante postulaciones electrónicas evaluadas por supervisores bajo criterios preestablecidos. El presente trabajo se centra en la predicción de la aceptación y acceso al PAF a través de las postulaciones electrónicas utilizando algoritmos de *machine learning*. La Inteligencia Artificial (IA) emerge como una tecnología disruptiva con potencial para optimizar procesos en diversos sectores, incluyendo el sector público y los programas sociales. Esta investigación busca explorar cómo el aprendizaje automático puede mejorar la eficiencia y el impacto de los programas sociales en Uruguay, específicamente en la identificación de beneficiarios.

## 2. DESCRIPCIÓN DEL CONJUNTO DE DATOS

Para llevar a cabo este estudio, se obtuvo una base de datos anonimizada del programa Uruguay Crece Contigo con información de derivaciones entre 2018 y 2024. El preprocesamiento de los datos incluyó la exploración de variables, la evaluación de la completitud y el tratamiento de valores faltantes. Se crearon dos bases de datos segmentadas: una con información de niños menores de cuatro años (BD 1) y otra con información de mujeres embarazadas (BD 2).

**Tabla 1. Descripción del conjunto de datos**

| Nro. | Tipología | Nro. de casos |
|---|---|---|
| 1 | Mujer embarazada, | 6677 |
| 2 | Niño/a menor de 4 años, | 7154 |
| 3 | Dos niños/as menores de 4 años, | 498 |
| 4 | Una mujer embarazada y un niño/a, | 60 |
| 5 | Una mujer embarazada y dos niños/as. | 840 |
| 99 | Sin dato | 207 |

**Origen y Tipo:** La fuente principal de datos fueron los casos históricos de derivación del programa Uruguay Crece Contigo del Ministerio de Desarrollo Social. Estos datos se obtuvieron mediante una solicitud formal y fueron anonimizados previamente para proteger la privacidad de los participantes.

**Período y Volumen:** El conjunto de datos incluyó 15,436 casos derivados entre los años 2018 y 2024.

**Enfoque:** Los casos se centraron en mujeres embarazadas y niños menores de cuatro años, que son la población objetivo del Programa de Acompañamiento Familiar (PAF).

**Tipologías de Derivación:** Los datos abarcan cinco posibles tipologías de derivación:

- Mujer embarazada
- Niño/a menor de 4 años
- Dos niños/as menores de 4 años
- Una mujer embarazada y un niño/a
- Una mujer embarazada y dos niños/as

**Segmentación de Datos:** Para el análisis, la base de datos se dividió en dos partes:

**BD 1 ("Niños"):** Conteniendo información sobre 7583 casos relacionados con niños, derivados de las tipologías 2, 3, 4 y 5.

**BD 2 ("Mujeres embarazadas"):** Conteniendo información sobre 7577 casos relacionados con mujeres embarazadas, derivados de las tipologías 1, 4 y 5.

**Variables:** Las bases de datos incluyeron diversas variables recopiladas a través del proceso de postulación electrónica. Algunas de estas variables fueron específicas para mujeres embarazadas o niños, mientras que otras fueron comunes a ambas. Ejemplos de variables incluyen:

- Fecha de derivación
- Origen de la derivación
- Departamento de residencia
- Información sobre transferencias monetarias recibidas por el hogar
- Fecha de nacimiento (para niños y mujeres embarazadas)
- Preguntas relacionadas con la salud (por ejemplo, controles prenatales, estado nutricional, presencia de anemia, vacunas para niños)
- Indicadores socioeconómicos (por ejemplo, condiciones de vivienda, seguridad alimentaria)
- Edad gestacional (para mujeres embarazadas)

**Datos Adicionales:** También se utilizó información sobre la cantidad de nacimientos por departamento, disponible públicamente en el sitio web de estadísticas vitales del Ministerio de Salud Pública. Estos datos se escalaron y se incorporaron tanto en la BD 1 como en la BD 2.

**Limitaciones:** El proceso de anonimización eliminó observaciones cualitativas potencialmente útiles y datos complementarios de registros administrativos. Además, hubo problemas con datos faltantes o incompletos, que variaron según la tipología de derivación. El diseño del formulario de solicitud también generó algunas respuestas irrelevantes según la edad del niño, lo que requirió limpieza de datos.

## 3. METODOLOGÍA

Se aplicaron técnicas de *feature engineering*, como la codificación *One-Hot Encoding* para variables categóricas y el escalado de variables numéricas. Se implementaron y compararon varios algoritmos de clasificación supervisada, incluyendo Regresión Logística, Árboles de Decisión, Random Forest, Redes Neuronales, Ensemble Voting, Stacking y, específicamente para BD 2, XG-Boost y LSTM. Se utilizaron métricas de evaluación como precisión (*precision*), exhaustividad (*recall*), F1-score, *accuracy* y la curva ROC con su área bajo la curva (AUC) para evaluar el rendimiento de los modelos. Se aplicaron estrategias para abordar el desbalance de clases como SMOTE y RUS, y se optimizaron los umbrales de decisión para mejorar el rendimiento de los modelos.

**Figura 1. Arquitectura del sistema**

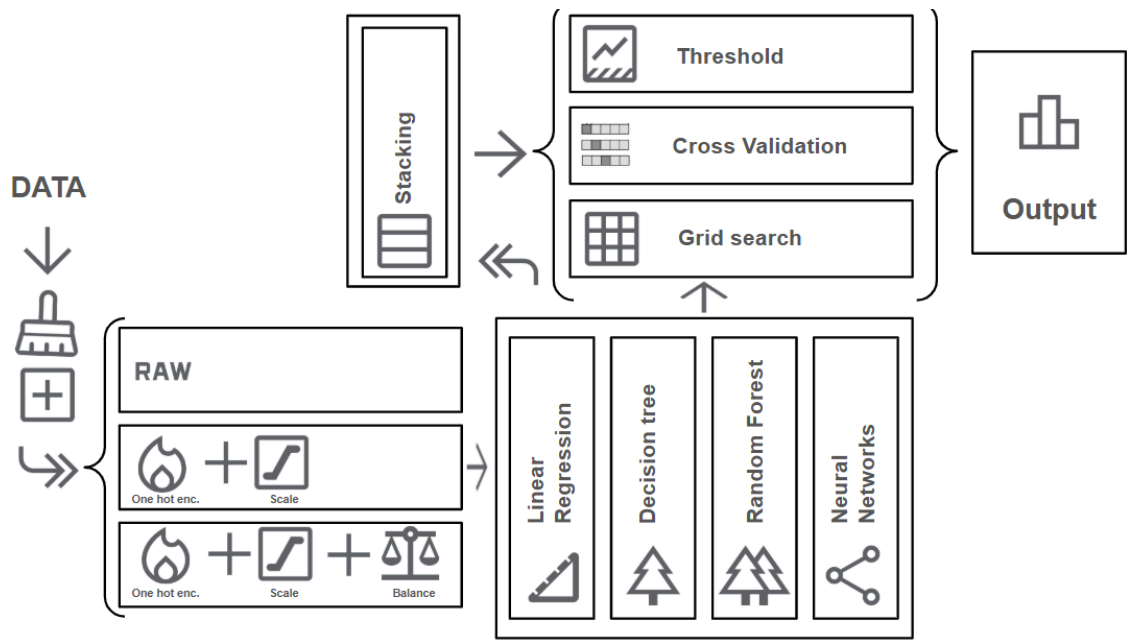

## 4. RESULTADOS

Los resultados obtenidos muestran que varios modelos de aprendizaje automático lograron un rendimiento aceptable en la predicción de la aceptación en el PAF.

En la BD 1, modelos como Regresión Logística, Random Forest y Stacking, con umbrales de decisión ajustados, demostraron una alta exhaustividad (*recall*), lo cual es importante para minimizar los falsos negativos.

En la BD 2, se exploraron diferentes configuraciones de los modelos, incluyendo el uso de SMOTE y RUS para balancear las clases. Modelos como XGBoost y las Redes Neuronales mostraron un buen equilibrio entre precisión y exhaustividad. El modelo de Stacking (NN - XGB - RF) también presentó un rendimiento destacado en la BD 2. La selección del umbral de predicción tuvo un impacto significativo en las métricas de evaluación, evidenciando la necesidad de ajustar este parámetro según los objetivos del programa (minimizar falsos positivos o falsos negativos)

**Tabla 1. Descripción del Base de Datos 1 y Base de atos 2**

**BD 1 (NIÑOS)**

| Algoritmo | Ajustes | Umbral | Precisión | Recall | F1-score |
|---|---|---|---|---|---|
| Regresión Logística | | 0.3 | 0.5704 | 0.9927 | 0.7245 |
| Árbol de Decisión | | 0.3 | 0.5522 | 0.931 | 0.6933 |
| Random Forest | | 0.3 | 0.5807 | 0.9428 | 0.7187 |
| Red Neuronal | | 0.3 | 0.6015 | 0.7109 | 0.6516 |
| Ensemble Voting | | 0.3 | 0.5974 | 0.909 | 0.721 |
| Stacking | | 0.3 | 0.5952 | 0.9428 | 0.7297 |

| BD 2 (MUJERES) | | | | | |
|---|---|---|---|---|---|
| XGBoost | GS+CV | | 0.72 | 0.96 | 0.82 |
| | SMOTE+GS+CV | 0.5 | 0.8 | 0.64 | 0.71 |
| | RUS + GS + CV | | 0,83 | 0,59 | 0,69 |
| Neural Network | (64, 32, 1) | 0.6 | 0,77 | 0,71 | 0,74 |
| | (64, 64, 32, 1) | 0.5 | 0,8 | 0,65 | 0,72 |
| | LSTM | 0.6 | 0,76 | 0,85 | 0,8 |
| Ensembled | Easy Ensembled Classifier | | 0,8 | 0,64 | 0,71 |
| | RUSBoostClassifier | | 0,82 | 0,53 | 0,64 |
| | RUSBoostClassifier + GS | | 0,71 | 0,97 | 0,82 |
| | Stacking(NN - XGB - RF) | 0.6 | 0,74 | 0,78 | 0,76 |

**Figura 2. Curva ROC para modelos y conjuntos**

**BD 1 (Niños)**  **BD 2 (Mujeres embarazadas)**

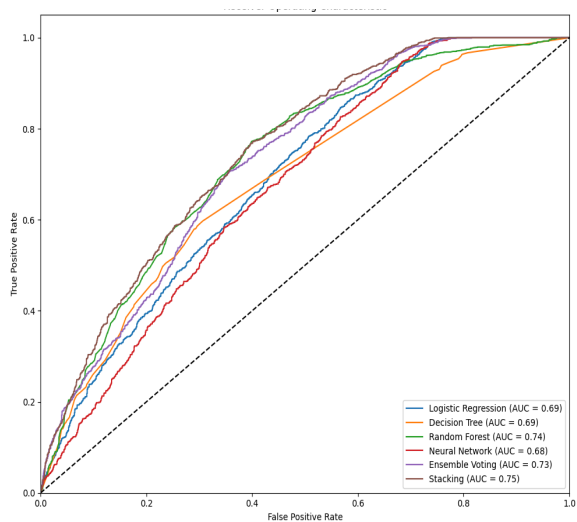
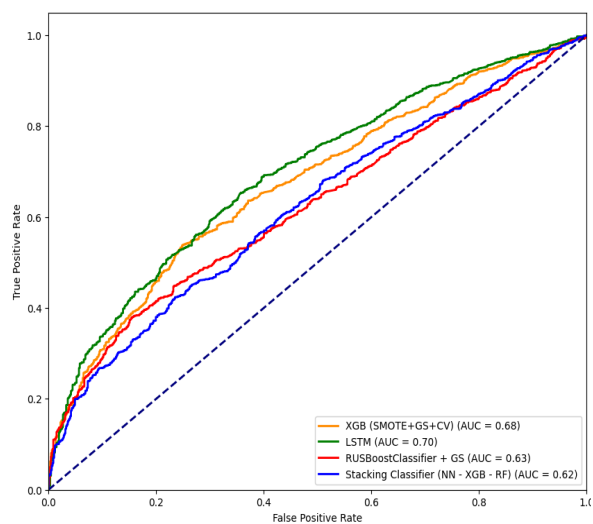

## 5. DISCUSIÓN Y CONCLUSIONES

El presente estudio demuestra el potencial del aprendizaje automático para optimizar el proceso de selección de beneficiarios en programas sociales. La implementación de un algoritmo predictivo puede agilizar la evaluación de postulaciones y permitir una asignación de recursos más eficiente. A pesar de los desafíos inherentes a los datos, como el desbalance y la incompletitud, los modelos implementados lograron resultados prometedores. Se subraya la importancia de considerar aspectos éticos, la transparencia en la toma de decisiones algorítmicas y la supervisión humana en el proceso. Hasta donde sabemos este trabajo representa la primera experiencia en aplicación de la IA vinculada a programas sociales en el sector público uruguayo.

**ANEXOS**

La versión completa se encuentra disponible en [este link](este link)